\documentstyle{mn}
\input{epsf}
\newif\ifAMStwofonts
\def\lesssim{\mathrel{\hbox{\rlap{\hbox{\lower4pt\hbox{$\sim$}}}\hbox{$<$}}}}

\def\gtrsim{\mathrel{\hbox{\rlap{\hbox{\lower4pt\hbox{$\sim$}}}\hbox{$>$}}}}

\def\msun{${\rm M_{\odot}}$}

\def\teff{$T_{\rm eff}$~}

\def\lll_lsun{$\log{\rm L/ L_{\odot}}$~}

\def\masa_msun{$M/ \rm M_{\odot}$~}

\def\m_mstar{$M/M_{*}$~}

\title{The Effects of Element  Diffusion on the Pulsational Properties
of Variable DA White Dwarf Stars}

\author[A.  H. C\'orsico, O.  G. Benvenuto,  L. G.  Althaus and  A. M.
Serenelli] {  A.  H.  C\'orsico\thanks{Fellow of  the Consejo Nacional
de   Investigaciones    Cient\'{\i}ficas   y   T\'ecnicas   (CONICET),
Argentina. Email: acorsico@fcaglp.fcaglp.unlp.edu.ar}
O.   G.   Benvenuto\thanks{Member  of  the  Carrera  del  Investigador
Cient\'{\i}fico, Comisi\'on de  Investigaciones Cient\'{\i}ficas de la
Provincia de Buenos Aires, Argentina.}
L.   G.    Althaus\thanks{Member  of  the   Carrera  del  Investigador
Cient\'{\i}fico y Tecnol\'ogico, CONICET.}
A.  M.  Serenelli\thanks{Fellow of  CONICET.}  \\ Facultad de Ciencias
Astron\'omicas  y Geof\'{\i}sicas, Universidad  Nacional de  La Plata,
Paseo del Bosque S/N, (1900) La Plata, Argentina}

\pagerange{\pageref{firstpage}--\pageref{lastpage}}

\pubyear{2001}

\begin{document}

\maketitle

\label{firstpage}

%----------------------------------------------------------------------
\begin{abstract}

We  explore the  effects  of element  diffusion  due to  gravitational
settling  and  thermal  and  chemical  diffusion  on  the  pulsational
properties  of DA white  dwarfs.  To  this end,  we employ  an updated
evolutionary code  coupled with a pulsational,  finite difference code
for  computing  the  linear,   non-radial  g-modes  in  the  adiabatic
approximation.  We follow  the evolution of a 0.55  \msun\ white dwarf
model  in  a  self-consistent  way  with  the  evolution  of  chemical
abundance distribution as given by time dependent diffusion processes.
Results  are  compared  with   the  standard  treatment  of  diffusive
equilibrium   in   the   trace  element   approximation.   Appreciable
differences are found between the two employed treatments. We conclude
that  time dependent  element  diffusion plays  an  important role  in
determining the whole oscillation  pattern and the temporal derivative
of the periods in DAV white dwarfs. 

In addition, we discuss the plausibility of the standard description
employed in accounting for diffusion in most of white dwarf 
asteroseismological studies. 

\end{abstract}

\begin{keywords}  stars:  evolution  -  stars: interiors - stars:
white dwarfs - stars: oscillations

\end{keywords}

%----------------------------------------------------------------------
\section{Introduction} \label{sec:intro}

Asteroseismology is a method to extract information about the internal
structure  and evolution  of  stars by  means  of the  study of  their
oscillatory pattern.   This technique, very sophisticated  in the case
of the Sun, has also undergone a strong development in other stars, in
particular the  pulsating white dwarfs (WDs) (for  reviews, see, e.g.,
Brown \& Gilliland 1994; Gautschy \& Saio 1995; 1996).

Pulsating  WDs  show  multi-periodic  luminosity variations  in  three
ranges  of   effective  temperatures  (\teff)   corresponding  to  the
currently called  DOV, DBV and DAV  (see, e.g., the  reviews by Winget
1988 and Kepler  \& Bradley 1995).  Of interest for  this work are the
DAVs (hydrogen-dominated atmospheres), or ZZ Ceti, that pulsate in the
instability  strip  corresponding  to  12500 K  $\gtrsim  T_{\rm  eff}
\gtrsim$ 10700 K.  The periodicities  in the light curves of pulsating
WDs  are naturally  explained in  terms of  non-radial g-modes  of low
harmonic degree ($\ell \leq  2$), driven by the ``$\kappa$ mechanism''
working in a partial ionization region near the stellar surface (Dolez
\& Vauclair  1981, Winget et  al.  1982).  Other  physically plausible
mechanism  for  overstability of  g-modes  in  ZZ  Ceti stars  is  the
``convective driving mechanism'' (see  Brickhill 1991 and Goldreich \&
Wu 1999  for details). The periods  ($P$) are found within  a range of
$100\ {\rm s} \leq P \leq  1200$ s and photometric amplitudes reach up
to 0.30 magnitudes.

Asteroseismology  of  WDs  has  recently  reached  important  success,
supplying  independent constraints  to several  structural quantities.
As a few  examples we mention the cases of DOV  PG 1159-035 (Winget et
al. 1991), DBV GD 358 (Bradley \& Winget 1994), and the DAVs G117-B15A
and R548 (Bradley 1996, 1998).

The main  observable feature in  WD pulsations is the  period pattern,
which can  be accurately measured.  Another important  quantity is the
temporal derivative of the  period ($\dot{P}$), that allows to measure
the  cooling time-scale  of  WDs  and to  provide  constraints on  the
chemical  composition  of the  core.   In  this  sense, the  star  DAV
G117-B15A is particularly noteworthy.  Its observed periods are 215.2,
271 and 304.4  s.  For the 215.2  s mode it has been  possible to find
its temporal derivative to be  $\dot{P}= (2.3 \pm 1.4) \times 10^{-15}
{\rm s\; s}^{-1}$ (Kepler et al. 2000).

As mentioned, one  of the main purposes of  asteroseismology of WDs is
to disentangle the observed periodic  signals in terms of the internal
structure and the evolution of  such objects.  In view of the detailed
available observations, it is  very important to study the pulsational
properties of DAVs  in the frame of evolutionary  models as physically
sound as possible. In this regard, most of existing calculations treat
the  chemical  profile  at  the hydrogen-helium  interface  (the  most
relevant one in the context of ZZ Ceti pulsations) on the basis of the
equilibrium  diffusion in the  trace element  approximation, hereafter
EDTE approximation,  (Tassoul, Fontaine \& Winget  1990, TFW; Bradley,
Winget  \&  Wood 1992;  Bradley  1996;  see  also Appendix).   In  the
treatment  used by  these  authors, the  chemical  profile may  change
solely as a result of changes in the state of ionization in the plasma
(see Appendix).   Thus, if the compositional  transition region occurs
at thermodynamical  conditions at which  the plasma is  fully ionized,
then such  a treatment predicts  fixed profiles. However, even  in the
case of  thick hydrogen envelopes  (when both hydrogen and  helium are
completely ionized  deep in the  star) element diffusion  modifies the
chemical abundance distribution within the star, and this is true even
during evolutionary  stages corresponding to  the ZZ Ceti  domain (see
Iben \& McDonald 1985, particularly their Fig.  4).

It is the aim of this work to perform new pulsation calculations in DA
WDs by relaxing both the trace element approximation and the diffusive
equilibrium  assumption.  To  this  end, we  carry out  time-dependent
diffusion   calculations   for    a   multicomponent   plasma   in   a
self-consistent  way   with  stellar  evolution.   Detailed  diffusion
calculations  consistent  with stellar  evolution  have recently  been
performed by McDonald,  Hernanz \& Jos\'e (1998) to  study the problem
of carbon pollution in cool WDs and by Althaus, Serenelli \& Benvenuto
(2001) to assess  the role  played by diffusion  in the  occurrence of
hydrogen  thermonuclear  flashes  in  low-mass, helium-core  WDs.   In
addition,  Dehner  \& Kawaler  (1995)  have considered  time-dependent
diffusion  in  evolving hot  WDs  in  the  interest of  exploring  the
possibility of an  evolutionary link between DO PG  1159 stars and the
much cooler DB  WDs.  In the context of pulsations,  the change in the
chemical composition  (particularly at the  hydrogen-helium interface)
induced by diffusion processes is  expected to affect the shape of the
Ledoux  term  $B$ and  hence  the  Brunt-V\"ais\"al\"a frequency  (see
Brassard  et al.   1991 for  a discussion  of the  calculation  of the
Brunt-V\"ais\"al\"a frequency in the context of WDs).

In order to gauge the actual importance of time-dependent diffusion in
the computation of theoretical $P$  and $\dot{P}$ we have to calculate
the  WD  cooling  considering  a full  evolutionary  code  considering
diffusion coupled to a pulsational code. To our knowledge, this is the
first time  that such kind of  calculation has been  undertaken in the
context of DA  WDs.  Note that in such kind  of treatment the internal
chemical profile is the  consequence of realistic evolutionary models.
Here,  we  present   calculations  of  linear,  adiabatic,  non-radial
pulsations  of  DAV  models with  a  mass  of  0.55 \msun\  (which  is
representative  of the mass  of G117-B15A).   In particular,  we shall
calculate  two evolutionary  evolutionary  sequences, one  considering
time-dependent element diffusion, and another  one in the frame of the
standard EDTE approximation.

The remainder of this paper is  organized as follows.  In Section 2 we
describe  our evolutionary-pulsational  computer code,  paying special
attention to the method for  simulating the diffusion of elements in a
time-dependent  approach.   Section 3  is  devoted  to presenting  the
calculations  we  performed. Finally,  in  Section  4  we discuss  our
results and make some concluding remarks.

%----------------------------------------------------------------------
\section{Our computer code}

\subsection{Evolutionary code and diffusion equations}

The evolutionary code we employed  is detailed in Althaus \& Benvenuto
(1997, 1998).  This code  is based on  a very detailed  and up-to-date
physical description such as  OPAL radiative (Iglesias \& Rogers 1996)
and molecular (Alexander \&  Ferguson 1994) opacities. The equation of
state is  an updated  version of that  of Magni \&  Mazzitelli (1979).
High-density conductive opacity and  neutrino emission rates are taken
from the  works of  Itoh and collaborators  (see Althaus  \& Benvenuto
1997 for details).  Also, a complete network of thermonuclear reaction
rates corresponding to the proton-proton chain and the CNO bi-cycle is
included.  Nuclear  reaction rates are from Caughlan  \& Fowler (1988)
and electron  screening is  treated as in  Wallace, Woosley  \& Weaver
(1982).

Gravitational settling,  and chemical and thermal  diffusion have been
fully taken  into account  following the treatment  for multicomponent
gases presented by  Burgers (1969).  Thus, we avoid  the trace element
approximation   usually  invoked  in   most  WD   studies.   Radiative
levitation,   which   is   important  for   determining   photospheric
composition of hot WDs (Fontaine \& Michaud 1979) has been neglected.

As  a  result of  gravity,  partial  pressure,  thermal gradients  and
induced  electric fields  (we  neglect stellar  rotation and  magnetic
fields) the  diffusion velocities  in a multicomponent  plasma satisfy
the  set of  $N-1$ independent  linear equations (Burgers 1969)

\begin{eqnarray}
{{dp_i} \over {dr}}-{{\rho  _i} \over \rho}{{dp} \over {dr}}-n_iZ_ieE=
\sum\limits_{j\ne   i}^{N}  {K_{ij}}\left({w_j-w_i}  \right)   \cr  +\
\sum\limits_{j\ne   i}^{N}  {K_{ij}\  z_{ij}}   {{m_j\  r_i\   -  m_i\
r_j}\over{m_i\ + m_j}},
\end{eqnarray}

\noindent and heat flow equation ($N$ equations)

\[{{5} \over {2}}n_i k_{B} \nabla T= - {{5} \over {2}}
\sum\limits_{j\ne   i}^{N}   {K_{ij}\   z_{ij}}   {{m_j}\over{m_i\   +
m_j}}\left({w_j-w_i} \right)  - {{2} \over  {5}}{K_{ii}}\ z_{ii}^{,,}\
r_i \]
\begin{eqnarray}
 -\sum\limits_{j\ne  i}^{N} {{K_{ij}}\over{(m_i\  +  m_j)^2}}\left( {3
m_i^2  + m_j^2  z_{ij}^{,}  + 0.8m_im_jz_{ij}^{,,}}  \right)\ r_i  \cr
{+\sum\limits_{j\ne i}^{N} {{K_{ij}m_im_j}\over{(m_i\ + m_j)^2}}\left(
{3 + z_{ij}^{,} - 0.8z_{ij}^{,,}} \right)\ r_j}.
\end{eqnarray}

In  these equations, $p_i$,  $\rho_i$, $n_i$,  $Z_i$ and  $m_i$ means,
respectively, the partial pressure, mass density, number density, mean
charge and mass for species $i$ ($N$ means the number of ionic species
plus electron). The quantities $T$ and $k_{B}$ are the temperature and
Boltzmann   constant.   The  unknown   variables  are   the  diffusion
velocities with respect to the centre of mass, $w_i$, and the residual
heat flows $r_i$  (for ions and electrons).  In  addition the electric
field $E$ has to be determined.  The resistance coefficients ($K_{ij},
z_{ij}, z_{ij}^{,}$ and $z_{ij}^{,,}$) are from Paquette et al (1986).

The set of  equations is completed by using the  conditions for no net
mass flow with respect to the center of mass

\begin{eqnarray}
\sum\limits_{i} {A_i} n_i w_i=0,
\end{eqnarray}

\noindent and no electrical current

\begin{eqnarray}
\sum\limits_{i} {Z_i} n_i w_i=0.
\end{eqnarray}

\noindent  In terms  of  the gradient  in  the number  density we  can
transform Eq. (1) to

\begin{eqnarray}
{{1}\over{n_i}}\left[\sum\limits_{j\ne  i}^{N} {K_{ij}}\left({w_i-w_j}
\right)+\ \sum\limits_{j\ne i}^{N} {K_{ij}\ z_{ij}} {{m_i\ r_j\ - m_j\
r_i}\over{m_i\  +  m_j}}\right] \cr  -\  Z_ieE=  \alpha_i  - k_{B}  T\
{{d\ln{n_i}}\over{dr}},
\end{eqnarray}

\noindent where

\begin{eqnarray}
\alpha_i = - A_i m_H g - k_{B} T\ {{d\ln{T}}\over{dr}},
\end{eqnarray}

\noindent being $A_i$, $m_H$, and $g$ the atomic mass number, hydrogen
atom mass and gravity, respectively.  Let us write the unknowns $w_i$,
$r_i$ and $E$ in terms of the gradient of ion densities in the form

\begin{eqnarray}
w_i    =   w_{i}^{gt}    -    \sum\limits_{ions   (j)}    \sigma_{ij}\
{{d\ln{n_j}}\over{dr}},
\end{eqnarray}

\noindent  where  $w_{i}^{gt}$ means  the  velocity  component due  to
gravitational settling and thermal  diffusion.  With Eqs.  (2) and (5)
together  with  (3)  and  (4)   we  can  easily  find  the  components
$w_{i}^{gt}$    and   $\sigma_{ij}$    by   matrix    inversions   (LU
decomposition). The evolution of the abundance distribution throughout
the star is  found by solving the continuity  equation. In particular,
we  follow the  evolution  of the  isotopes  $^1$H, $^4$He,  $^{12}$C,
$^{14}$N and  $^{16}$O.  To calculate the dependence  of the structure
of  our WD  models on  the evolving  abundances  self-consistenly, the
diffusion equations have been coupled to the evolutionary code.

\subsection{The pulsational code}

In order to  compute the g-modes of the WD models  we have coupled our
evolutionary  code to  our  new, finite  difference, pulsational  code
described in C\'orsico \& Benvenuto (2002), which solves the equations
for linear, adiabatic, non-radial pulsations (Unno et al. 1989).

We  describe now how  these codes  work together.   To begin  with, an
interval  in $P$ and  $T_{\rm eff}$  ($T_{\rm eff}$-strip)  is chosen.
The evolutionary code computes the model cooling until the hot edge of
the  $T_{\rm eff}$-strip  is  reached.  Then,  the  program calls  the
pulsation routine beginning the scan for modes.  When a mode is found,
the  code  generates  an  approximate solution  which  is  iteratively
improved to convergence and  stored.  This procedure is repeated until
the period interval is covered.  Then, the evolutionary code generates
the next stellar  model and calls pulsation routines  again.  Now, the
previously  stored modes  are taken  as initial  approximation  to the
modes of  the new stellar model  and iterated to  convergence.  Such a
procedure is automatically repeated for all evolutionary models inside
the  $T_{\rm eff}$-strip. The  computational strategy  described above
has  been successfully  applied  in fitting  the  observed periods  of
G117-B15A to  impose constraints on  the mass of axions  (C\'orsico et
al. 2001a) and in computing the period structure of low mass, helium 
WDs (C\'orsico \& Benvenuto 2002).

We  have tested  our pulsational  code  with two  carbon-oxygen DA  WD
models  of 0.5  \msun\  and 0.85  \msun,  the structure  of which  was
computed with the WDEC  evolutionary code.  The vibrational properties
of such  models were  previously analyzed by  Bradley (1996).   In the
interests of a detailed comparison,  we have considered a large amount
of modes  and we found  that the differences  between the two  sets of
modes remain below $\approx 0.1 \%$.

%----------------------------------------------------------------------
\section{Computations} \label{sec:compu}

We  have evolved  a 0.55  \msun\ WD  model with  an internal  carbon -
oxygen chemical profile corresponding to that calculated by Salaris et
al. (1997).   Such a model has  hydrogen and helium  mass fractions of
$M_{\rm   H}/M_{*}=   10^{-4}$   and   $M_{\rm   He}/M_{*}=   10^{-2}$
respectively.  These  values are  in good agreement  with evolutionary
predictions and are also very similar to those found by Bradley (1998)
for the case of G117-B15A.  The internal chemical profile of our model
is shown in Fig. 1.  It is  important to mention that at the bottom of
the  hydrogen envelope  of our  model, hydrogen  and helium  are fully
ionized and  this is so  throughout the entire evolutionary  stages we
study  in the  present  paper. Thus,  the  chemical abundance  profile
predicted  by   the  trace  element  approach   remains  fixed  during
evolution.  In  computing radiative opacities, we  have assumed $Z=0$.
We have treated  convective transport in the frame  of the ML3 version
of the mixing length theory.  The ML3 prescription, characterized by a
high convective efficiency, assumes the  mixing length to be two times
the local pressure scale height (see Tassoul et al. 1990).

A realistic starting model for our evolutionary sequences was obtained
by artificially brightening an initial WD configuration (see Benvenuto
\& Althaus  1998) up to  \lll_lsun= 2.  Such  a procedure is  known to
produce  an  initial sequence  of  some  unphysical  models, but  then
relaxes to the correct cooling sequence (see Althaus \& Benvenuto 2000
for further discussion) far before reaching the DAV instability strip.
Since  then on  element  diffusion is  incorporated.   When the  model
reaches   \teff=   14000   K   we  start   pulsational   calculations.
Specifically we  have calculated  dipolar ($\ell=1$) modes  (which are
usually encountered in ZZ Ceti  light curves) with radial orders $k=1,
\cdots, 21$  which cover  a period interval  $100\ {\rm s}  \lesssim P
\lesssim 1000$ s.   Calculations are stopped at \teff=  10000 K, thus,
the  \teff-strip amply  embraces the  observed DAV  instability strip.
For  the modes  we  have found  to  fulfill such  conditions, we  have
computed  periods  and  eigenfunctions.   For computing  the  Brunt  -
V\"ais\"al\"a frequency, we have employed the appropriate prescription
for degenerate models, given in Brassard et al.  (1991).  After period
assessment we compute $\dot{P}$ by numerical differentiation.

%----------------------------------------------------------------------
\section{Results and Implications} \label{sec:res_imp}

We begin  by examining Fig. 2, in  which we show the  evolution of the
chemical  profile for  helium  resulting from  time dependent  element
diffusion  together with the  chemical profile  arising from  the EDTE
approximation.   Note that  in the  latter case,  the  profile remains
unchanged   throughout  the  evolution   because,  as   we  mentioned,
ionization  is   complete  at   such  deep  layers   \footnote{In  the
calculations presented  here, the EDTE approximation  has been applied
only to the  hydrogen-helium interface, which is the  most relevant in
the  context of DA  WD pulsations.}.   By contrast,  in the  case with
time-dependent diffusion, the  chemical abundance distribution evolves
appreciably  during the ZZ  Ceti evolutionary  stages.  Note  that the
shape  of the  profile in  both treatments  turns out  to  be markedly
different particularly at  the centre of the transition.   As we shall
show below,  this will  have an appreciable  influence on the  $P$ and
$\dot{P}$ values for some of the modes.

In Fig.   3, we  depict the  resulting Ledoux term  (panel A)  and the
squared  Brunt-V\"ais\"al\"a  frequency  (panel  B)  at  two  selected
$T_{\rm  eff}$ values.  Let  us remind  the reader  that the  term $B$
depends not only on the shape  of the chemical profile but also on the
thermal  and mechanical  structure of  the star  (see Brassard  et al.
1991).  Thus, even  in the  EDTE approach  the term  $B$  changes with
cooling.  Because  of the fact that  in the Ledoux  term there appears
the derivative of the chemical profile  (see Eq.  35 of TFW), a slight
change in the slope of the hydrogen-helium interface translates into a
noticeable change  in $B$.   Thus, it is  not surprising the  $B$ term
exhibits a sharp  peak in the case of the  trace element treatment, in
contrast with the more  physical treatment as given by non-equilibrium
diffusion.    In   turn,   this    feature   is   reflected   in   the
Brunt-V\"ais\"al\"a frequency.   As it has been  exhaustively shown by
Brassard  et  al.   (1992ab),  the shape  of  the  Brunt-V\"ais\"al\"a
frequency at  the chemical interfaces plays  a key role  in fixing the
structure  of the  period pattern  (e.g.,  mode trapping)  of ZZ  Ceti
stars.

Now let  us turn our attention  to the computed  pulsational modes. In
Figs. 4  and 5 we show  $P$ and $\dot{P}$ corresponding  to modes with
$\ell=1,  k=1,  \cdots,  6$  for models  with  time-dependent  element
diffusion  and  with the  EDTE  approximation  at the  hydrogen-helium
interface as a function of  $T_{\rm eff}$.  From a close inspection of
these figures  it can be  realized that the effects  of time-dependent
element  diffusion are  indeed  non-negligible in  $P$ and  $\dot{P}$,
although for some  modes results are very similar.  We want to mention
that the  same trend has  been found in  modes of higher  radial order
(not shown  here for  brevity).  Note that  for the modes  analyzed in
these  figures  the  greatest  relative  differences  encountered  are
$\approx 20\%$ for $\dot{P}$ and $\approx 5\%$ for $P$.

It is worth  mentioning that the differences cited  above arise mainly
from the very different shape  of the interface profile resulting from
the two treatments of diffusion investigated here.  In particular, the
differences in  the helium profile  for $X_{\rm He} \gtrsim  0.5$ (see
Fig. 2)  in these treatments are  the main reason why  the periods and
period  derivatives become different.  In addition,  there is  a small
contribution  to the  differences  in  $P$ and  $\dot{P}$  due to  the
evolution of  the profiles of  each chemical interface in  response to
non-equilibrium diffusion.

Thus,  as  models  with  time  dependent element  diffusion  are  more
physically  plausible, these  should  be taken  into  account when  an
asteroseismological fit  to observed  periods is performed.   Also, as
$\dot{P}$ is modified, this approach should also be taken into account
at using observed $\dot{P}$ values  to infer the composition of the WD
core.

Before closing  the paper, we would  like to discuss at  some length a
major issue raised  by our referee.  Indeed, in  his (her) report, our
referee asked us  to look for the underlying  physical reasons for the
differences between  the standard treatment of  EDTE approximation and
our  full  treatment  of  time  dependent element  diffusion.   To  be
specific, the referee asked us whether such differences are due mostly
to  the   relaxation  of  the  trace  element   approximation  or  the
equilibrium hypothesis.

In order to find the answer,  we decided to perform a simple numerical
experiment: we simulate the  equilibrium diffusion conditions with our
full code. Equilibrium diffusion would  be a good approximation if the
diffusion  timescale  were much  shorter  than  the evolutionary  one.
Then,  in order  to simulate  this  situation, we  simply assumed  the
diffusion   time   step  to   be   several   times  the   evolutionary
one\footnote{Here,  because   of  numerical  reasons,   we  assumed  a
diffusive time step  a hundred times the evolutionary  one.}.  This is
equivalent to assuming that the whole diffusive process occurs several
times faster.   We computed the evolution  of our WD  model under this
hypothesis. The result  was that {\it the WD did  not evolve along the
cooling  branch but  instead  suffered from  a hydrogen  thermonuclear
flash}.   Physically,  the  reason  for  this behaviour  is  that,  if
diffusion were plenty of time  to evolve to equilibrium profiles, then
the tail  of the hydrogen  profile would be  able to reach  hot enough
layers to  be ignited in a  flash fashion.  The fact  that at imposing
equilibrium  diffusion  conditions   the  star  undergoes  a  hydrogen
thermonuclear  flash,  while  in  the detailed,  self-consistent  time
dependent diffusive treatment the star cools down quiescently, clearly
shows the incorrectness of the  hypothesis of equilibrium.  In view of
this, we  are forced to  conclude that there  is no way other  than to
abandon the idea that the shape  of the internal profiles in the WD is
determined  by equilibrium  diffusion.   This conclusion  is valid  at
least  for  massive  hydrogen  envelopes.   We  think  that  the  only
physically sound  way to  compute such profiles,  a key  ingredient in
asteroseismological  studies, is to  calculate the  WD evolution  in a
self-consistent way with time  dependent element diffusion and nuclear
burning.

In addition, some  words are in order about  the standard treatment of
equilibrium  diffusion, which  is based  on  the work  of Arcoragi  \&
Fontaine  (1980).   Recent  asteroseismological  studies  of  DAV  WDs
(Clemens 1994, Bradley 1998, 2001) seem to favour large values for the
thickness   of  the  hydrogen   envelope  ($M_{\rm   H}/M_{*}  \approx
10^{-4}$).   At  the thermodynamical  conditions  relevant  to DAV  WD
models, we find  that most of the hydrogen-helium  interface occurs at
degenerate  conditions.   Because  the  Arcoragi  \&  Fontaine  (1980)
equations  are valid  for {\it  non-degenerate} conditions,  we should
remark that  this treatment cannot be  applied to the  modeling of the
hydrogen-helium interfaces in DA WDs with massive hydrogen envelopes.

We  should  also  remark  that,  in  the case  of  the  trace  element
approximation,  we have  found that  the object  does not  undergo any
thermonuclear flash.   This is another artifact  of the approximation,
due to the {\it ad hoc}  truncation of the profile at some low density
(see Appendix).   Indeed, in our numerical experiments,  we have found
that  the stellar  model  evolves along  the  WD cooling  track if  we
truncate the hydrogen profile at  $X_{\rm H}= 10^{-3}$. However, if we
allow  the equilibrium hydrogen  profile to  extend to  slightly lower
abundances  (e.g.,  $X_{\rm  H}=  10^{-4}$) the  model  experiences  a
thermonuclear flash!

The results presented in this paper indicate that a more extensive and
systematic  exploration of  asteroseismology of  DAV in  the  frame of
detailed  evolutionary  models   considering  time  dependent  element
diffusion is  worth being done and  it will be the  subject of further
papers.  While the  present paper  was in  process of  reviewing, some
interesting results  about the effects  of diffusion on  mode trapping
have been presented in C\'orsico et al. (2001b).

%----------------------------------------------------------------------
\section{Acknowledgments} \label{sec:acknow}

The authors would like to acknowledge Paul Bradley for his kindness in
providing us  with models  of carbon-oxygen DA  WDs together  with his
pulsation calculations.   This allowed us to  test our code  and to be
confident  with the  results our  pulsational code  produces. It  is a
pleasure to thank our referee, whose comments and suggestions strongly
improved the original version of this paper.
%-------------------------------------------------------------------

\bsp

\label{lastpage}

%----------------------------------------------------------------------

\section*{Appendix}

For the sake of completeness,  we describe here the equations employed
to  include diffusion  processes in  most of  WD  pulsational studies.
This approximation is based on the work of Arcoragi \& Fontaine (1980)
(see also Tassoul  et al.  1990).  Here we  limit ourselves to comment
on the most important aspects involved in this treatment.

To begin  with, Arcoragi  \& Fontaine (1980)  assume a  stellar plasma
made up of  two-ionic species with average charge  $Z_1$ and $Z_2$ and
atomic  weight $A_1$  and $A_2$.   In addition,  thermal  diffusion is
neglected and an  ideal gas equation of state  is considered under the
assumption  that  the plasma  is  sufficiently  diluted.  Under  these
approximations, the diffusion velocity $w_{12}$ reads    

\begin{eqnarray}
w_{12} & = & D_{12}(1+\gamma) \bigg[-\frac{\partial \ln{c_2}}{\partial
r} + \frac{A_2-A_1}{A_1+\gamma A_2} \frac{\partial \ln{p}}{\partial r}
+  \nonumber\\ &  & \frac{A_2  Z_1 -A_1  Z_2}{A_1+\gamma  A_2} \frac{e
E}{k_{B} T} \bigg].
\label{eq:w}
\end{eqnarray}

\noindent $D_{12}$ is the diffusion coefficient, and $c_i$, the number
concentration of ions of species $i$, is defined by

\begin{equation} 
c_i \equiv \frac{n_i}{n_1+n_2} = \frac{p_i}{p_1+p_2} \label{eq:ci}
\end{equation}

\noindent being  $p_i$ the  partial pressure.  $E$ is  the electric
field, given by

\begin{equation}
e E= m_p g \frac{A_1 Z_1 + A_2 Z_2 \gamma}{Z_1 \left(Z_1+1\right)+ Z_2
\left(Z_2+1\right)\gamma},
\end{equation}

\noindent and $\gamma$ is defined as

\begin{equation}
\gamma \equiv \frac{n_2}{n_1} = \frac{p_2}{p_1}= \frac{c_2}{c_1}.
\label{eq:gamma}
\end{equation}

\noindent The remainder of the symbols have the usual meaning.  Notice
that  Eqs.   (\ref{eq:ci})  and   (\ref{eq:gamma})  are  valid  in  an
isothermal medium, i.e., we are neglecting temperature gradients.

Now,  we impose equilibrium  diffusion, by  assuming $w_{12}=  0$, and
from  Eqs.    (\ref{eq:w}  -  \ref{eq:gamma})  we   get  the  ordinary
differential equation for the  equilibrium profile (Eq. A5 of Arcoragi
\&  Fontaine 1980). In  the trace  element approximation  ($\gamma \ll
1$), for the specie 2 considered as a trace, we get

\begin{equation} \label{c2}
\frac{\partial \ln  c_2}{\partial r} = \alpha_2  \frac{\partial \ln q}
{\partial r},
\end{equation}

\noindent where

\begin{equation} 
\alpha_2= \frac{A_2}{A_1} \left(1 + Z_1\right) - Z_2 - 1.
\end{equation}

For  the purpose  of application,  Tassoul et  al.  (1990)  divide the
transition zone  into two parts:  an upper one  in which element  1 is
dominant and element 2 is a trace,  and a lower one in which the roles
of  the respective  elements is  reversed.  For  the upper  region the
abundance profile of  element 2 considered as a trace  is given by Eq.
(\ref{c2}) and for the lower part of the transition zone the abundance
of element 1 (considered as a trace) is given by

\begin{equation} \label{c1}
\frac{\partial \ln  c_1}{\partial r} = \alpha_1  \frac{\partial \ln q}
{\partial r},
\end{equation}

\noindent where

\begin{equation} 
\alpha_1= \frac{A_1}{A_2} \left(1 + Z_2\right) - Z_1 - 1.
\end{equation}
 
\noindent $q$ is the mass fraction ($1 - M_r / M_*$).  The integration
of  Eqs.  (\ref{c2})  and (\ref{c1})  gives the  equilibrium abundance
profiles:

\begin{equation} 
c_2= k_2 q^{\alpha_2} \\ \textrm{(upper region of interface)}
\end{equation}

\noindent and

\begin{equation} 
c_1= k_1 q^{\alpha_1} \\ \textrm{(lower region of interface)}
\end{equation}

By invoking  the condition  of continuity in  the middle point  of the
inteface, we obtain the relation

\begin{equation} 
k_2 q_m^{\alpha_2}= k_1 q_m^{\alpha_1}= \frac{1}{2},
\end{equation}

\noindent where $q_m$ is the mass fraction where the abundances of two
species are  equal; the  $q_m$ value is  obtained by forcing  the mass
conservation of element  1.  Thus, in the case  of the hydrogen-helium
transition region,  the outer mass  fraction of hydrogen  ($q_{\rm H}=
M_{\rm H} /M_*$) is employed  for computing $q_m$.  Note that possible
changes in the equilibrium profiles result only from slight changes in
the ionization states of the elements present at the interfaces, i.e.,
variations  in the  exponents  $\alpha_1$ and  $\alpha_2$ (Tassoul  et
al. 1990).

To  implement  this  approach   to  modeling  of  the  hydrogen-helium
transition zone,  it is necessary to  set small abundances  to zero in
order to avoid having a tail  of hydrogen in regions deep enough where
carbon  is abundant  (for this  case  we should  generalize the  above
treatment for three  species). Moreover, if we do  not do so, hydrogen
would be present  at layers hot enough to force the  star to undergo a
thermonuclear flash.  This is  the case, at  least for  thick hydrogen
envelopes  like those we  have treated  here which,  in turn,  are the
favoured  ones by  current asteroseismological  studies (see  the main
text).

%----------------------------------------------------------------------

% figure 1
\begin{figure*}
\vskip 4.2cm \epsfysize=550pt \epsfbox{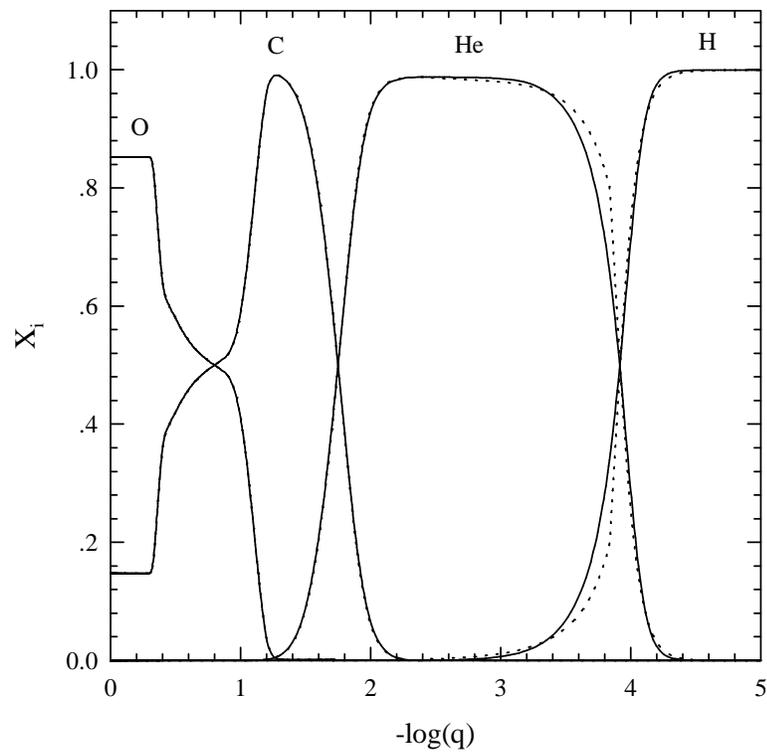}
\vskip -20mm
\caption{The   internal   chemical  profiles   of   the  0.55   \msun\
carbon-oxygen WD model  for hydrogen, helium, carbon and  oxygen at an
effective   temperature  of  14000K.    In  the   case  of   the  EDTE
approximation,   the  fixed   profiles  are   represented   by  dotted
lines. Profiles  for models in which time  dependent element diffusion
has been considered  are represented by solid lines.  $q$ is the outer
mass fraction defined by $q= 1- M_r/M_*$.}
\end{figure*}

% figure 2
\begin{figure*}
\epsfysize=550pt \epsfbox{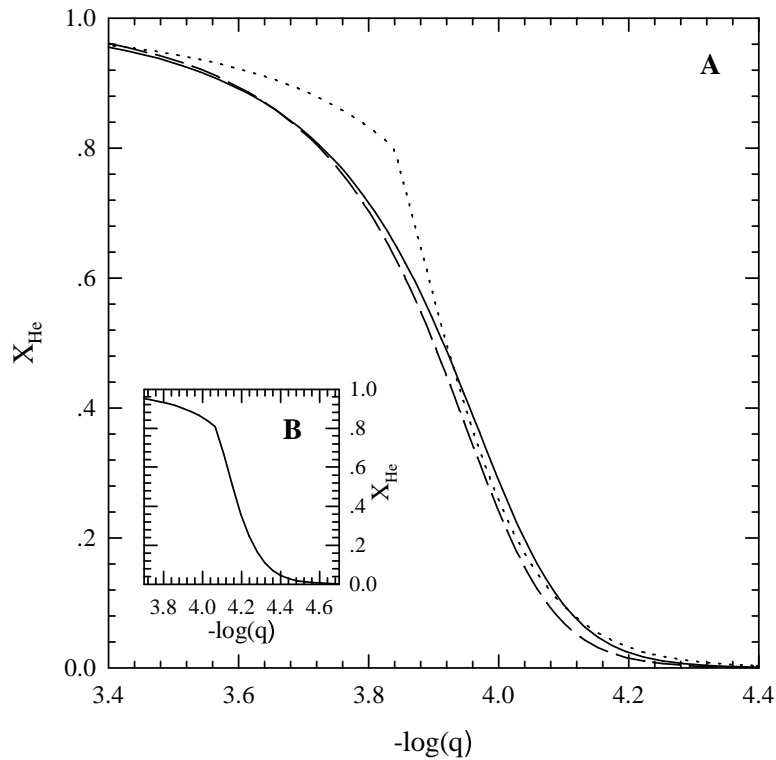}
\caption{Panel A: The internal chemical profile for helium in terms of
the outer  mass fraction $q$ at the  hydrogen-helium interface.  Solid
line  corresponds to  a  model at  a  temperature of  14000K in  which
time-dependent diffusion  is considered.   Long dashed line  means the
same  treatment  but  for  a  model at  10000K.  Finally  dotted  line
corresponds  to  the  EDTE  approximation  prediction.  Panel  B:  The
chemical  profile  for  helium  as  given by  the  EDTE  approximation
according  to  a  0.50  \msun  model calculated  by  Bradley  (private
communication). Note that the shape of the profile in both models with
diffusive equilibrium is the same.}
\end{figure*}

% figure 3
\begin{figure*}
\epsfysize=480pt \epsfbox{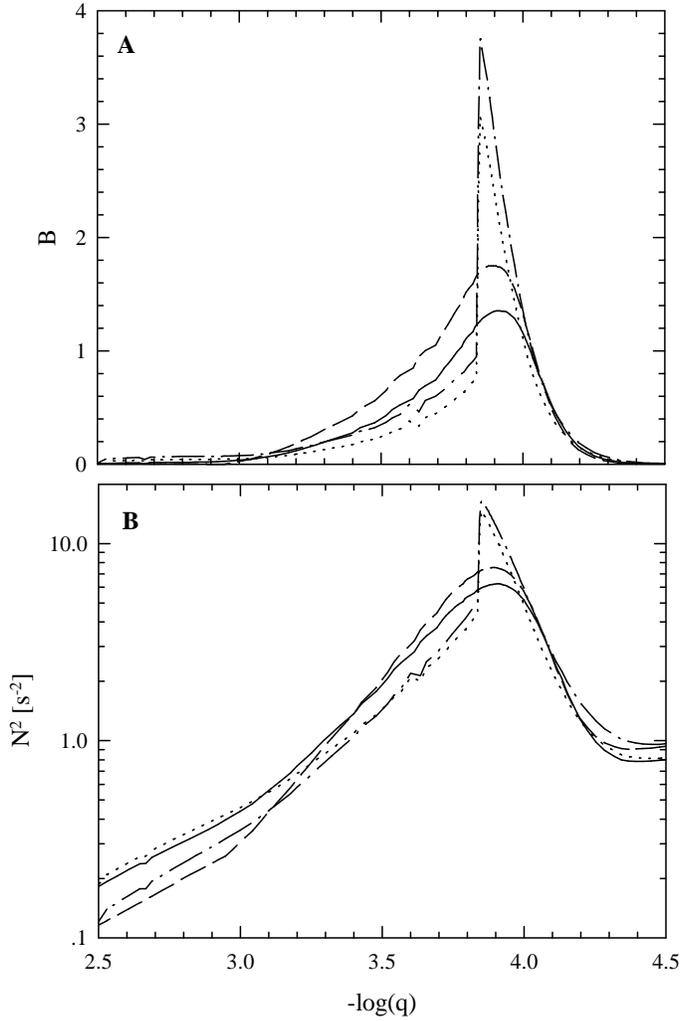}
\caption{In the  upper panel it  is shown the  Ledoux term $B$  at the
hydrogen-helium interface  in the case of  time-dependent diffusion at
the $T_{\rm  eff}$ values of 14000K  and 10000K as given  by solid and
long  dashed   lines,  respectively.   Dotted   and  dot-dashed  lines
correspond  to   the  same  temperature   values  but  for   the  EDTE
approximation.    In    the   lower   panel   the    square   of   the
Brunt-V\"ais\"al\"a frequency is shown  for the same cases analyzed in
panel A. For details, see text.}
\end{figure*}

% figure 4
\begin{figure*}
\epsfysize=480pt 
\epsfbox{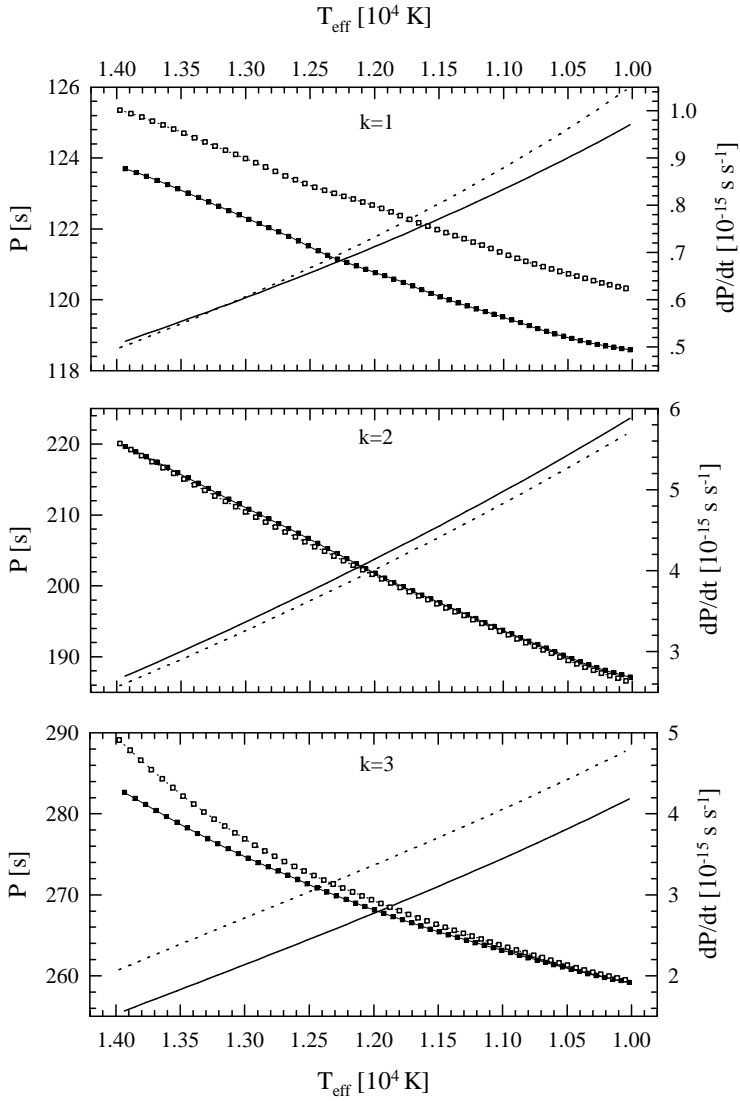}
\caption{Period and  period derivative for  $\ell=1, k=1, 2,  3$ modes
for  a  0.55 \msun\  carbon-oxygen  WD model  in  at  a $T_{\rm  eff}$
interval containing  the DAV  instability strip.  Solid  lines (filled
squares)   correspond  to   periods   (period  derivatives)   computed
considering  non-equilibrium  diffusion   while  dotted  lines  (empty
squares) depict periods (period derivatives) computed according to the
EDTE approximation at the hydrogen-helium interface. For discussion of
the results, see text.}
\end{figure*}

% figure 5
\begin{figure*}
\epsfysize=480pt 
\epsfbox{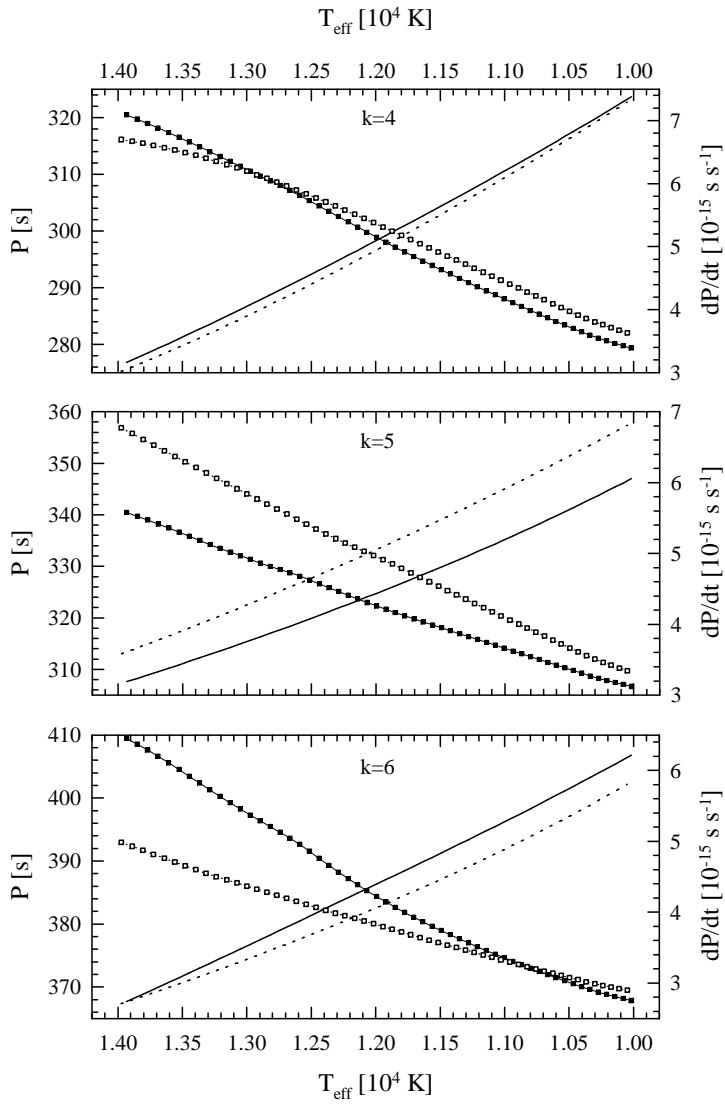}
\caption{Same as figure 4 but for modes
with $\ell=1, k=4, 5, 6$.} 
\end{figure*}

\end{document}